%% file: main_bis.tex
\documentclass{llncs}

\usepackage{etex}
\usepackage{graphicx}
\usepackage{multirow}
\usepackage{algorithm}
\usepackage{subcaption}
\usepackage{url}
\usepackage{amsmath}
\usepackage{balance}
\usepackage{wasysym}
\usepackage{longtable}
\usepackage{latexsym}
\usepackage{amssymb}
\usepackage{amstext}
\usepackage{xspace}
\usepackage{hyperref}
\usepackage{todonotes}
\usepackage{tabularx}
\usepackage{paralist}
\usepackage{rotating}
\usepackage{graphicx}
\usepackage{lipsum}%
\usepackage{wrapfig}
\usepackage{stmaryrd}
\usepackage{oz}
\usepackage{multirow}
\usepackage{mathtools}

\usepackage{amsopn}
\usepackage{fancybox}
\usepackage{helvet}
\usepackage{listings}
\usepackage{pifont}
\usepackage{rotate}
\usepackage{amssymb,amsmath}
\usepackage{url}
\usepackage{listings}
\usepackage{tocloft}
\usepackage{fancyhdr}
\usepackage{booktabs}
\usepackage{listings}

\usepackage{algorithmic}
\usepackage[ruled,linesnumbered,algo2e]{algorithm2e}

\graphicspath{{figures/}}

\usepackage{todowarning}

\begin{document}

\title{A Comparative Evaluation of Log-Based Process Performance Analysis Techniques}

\author{Fredrik Milani \and Fabrizio M. Maggi}
\institute{
    University of Tartu, Estonia		
	\\
    \email{milani@ut.ee, f.m.maggi@ut.ee}
}

\makeatletter
\makeatother

\maketitle
\begin{abstract}
\vspace{-0.3cm}
\input{abstract}
\end{abstract}

\setcounter{footnote}{0}

\section{Introduction}
\label{sec:intro}
\input{intro}
\input{review}
\input{results}

\input{framework}

\input{conclusion}

\section*{Acknowledgments}
This project and research is supported by Archimedes Foundation and GoSwift O\"{U} under the Framework of Support for Applied Research in Smart Specialization Growth Areas.








\bibliographystyle{splncs}
\bibliography{theBibliography}

\end{document}

%% file: abstract.tex

Process mining has gained traction over the past decade and an impressive body of research has resulted in the introduction of a variety of process mining approaches measuring process performance. Having this set of techniques available, organizations might find it difficult to identify which approach is best suited considering context, performance indicator, and data availability. In light of this challenge, this paper aims at introducing a framework for categorizing and selecting performance analysis approaches based on existing research. We start from a systematic literature review for identifying the existing works discussing how to measure process performance based on information retrieved from event logs. Then, the proposed framework is built starting from the information retrieved from these studies taking into consideration different aspects of performance analysis. 

%% file: intro.tex
Businesses are at a turning point where they have to incorporate digitalization or fade away. Digital technologies continue to set their transformative marks on virtually all industry domains and have allowed the expansion of businesses to markets previously inaccessible. The forces of innovation and creativity have enabled young businesses to challenge incumbents in practically every sector. However, one thing has not changed. Businesses will always seek to improve their processes because ``every good process eventually becomes a bad process'' \cite{hammer2015business}. This is even more relevant in a fast-changing digital era.

The first step to process improvement is to understand where processes can be improved. In the past, given the lack of data availability and high cost of data processing, performance analysis methods identified improvement opportunities based on manual analysis, and at times combined with random sampling (e.g., six sigma \cite{pande2000six}). Relying on such manually driven methods, process analysts assessed the performance of processes so to find opportunities for improvement. Today, much of the data is captured digitally and, over the past decade, analysis of large sets of data has improved remarkably. No longer are businesses restricted to select the most prioritized processes, limit the scope, or confine the selection of data due to limitations of time-consuming analysis or tools. In addition, the accessibility to open source tools has never been easier, in particular for data driven analysis of business processes. Therefore, in a digital era, businesses cannot hope to survive with manually driven methods. The process analysis must also be digitally transformed by tapping into data-driven analysis methods.

One group of techniques for data driven performance analysis uses event logs of processes to assess performance. Indeed, nowadays, business processes are often supported by IT systems that log their execution. For instance, an order-to-invoice process might include activities such as \textsc{register}, \textsc{validate}, \textsc{approve}, \textsc{fill order} and \textsc{send invoice}. Each order has a unique id and every activity is recorded in the event log with information about the time when it was executed (timestamp) and other additional data such as the resource (person or device) that executed the activity. As such, the process is inherently captured in the log. With process mining techniques \cite{van2016process}, the performance of such processes can be assessed and analyzed in great detail based on event logs.

The body of research and tools within the field of process mining has grown significantly during this decade. However, the availability of tools and approaches developed for specific aspects of process performance does not make it easier for businesses to employ them. In fact, it poses a challenge. There is no way for businesses to easily get an overview of what performance indicators can be measured, what input data is required for such analysis, or what industry specific implementations are available. In light of this context, we propose a framework for the selection of log-based performance analysis techniques. We do so by conducting a systematic literature review to identify the body of existing work. We analyze the results and focus on  identifying existing process performance indicators, required input data, and approaches available. Based on the results, we build a framework for the selection of suitable performance analysis approaches.

The structure of this paper is as follows. Section 2 summarizes the research protocol for the systematic survey. In Section 3, the research questions are discussed and the framework is presented in Section 4. Finally, Section 5 concludes the paper.

%% file: review.tex
\section{Systematic Literature Review}\label{sec:background}
In this section, we summarize how the systematic literature review was conducted. The review protocol specifies research questions, search protocol, inclusion and exclusion criteria, and data extraction. The review protocol predominantly follows the guidelines provided by Kitchenham~\cite{kitchenham2004procedures}.
The objectives of this paper are to review the current academic research on performance analysis techniques based on logs and build a framework for categorizing them. To this end, the overarching research question of ``what is the body of relevant academic research within the field of process performance analysis?'' has been decomposed into three sub-questions:
\begin{itemize}
  \item \textbf{RQ1}: \emph{What aspects of process performance do existing techniques consider?} This research question aims at identifying the aspects that can be measured in regards to process performance.
  \item \textbf{RQ2}:\emph{ What input data is required for measuring process performance?} For performance analysis, it is important that the ``right'' set of data is captured. To this end, it is important do understand what kind of data is required as input for process performance analysis techniques.
  \item \textbf{RQ3}: \emph{What are the main approaches/algorithms and tools available for analysis of process performance?} The final research question aims at capturing the various methods that can be applied for performance analysis.
\end{itemize}


To find relevant studies, we sought studies within the domain of ``process mining''. However, as process mining covers many aspects of business process analysis, such as process discovery \cite{van2011process}, we included ``performance'' to focus the search. The boolean search string (``process mining'' AND ``performance'') was used. The search was applied to Web of Science (WoS) and Scopus databases. These electronic databases were selected as they are the primary databases for computer science research. The search was conducted in January 2018 and resulted in a total of 330 studies from Scopus and 194 from WoS. After having removed duplicates, 349 studies remained. The first filtering was aimed at removing studies that were clearly out of scope (based on title), shorter than 6 pages, not accessible, or not written in English. The abstract and introduction of the remaining 101 studies were examined. Peer reviewed studies introducing or extending an existing approach for performance analysis, or directly dealing with performance of a process were included. After this filtering, the list of studies was reduced to 48. These studies were examined in full following the inclusion criteria of being within the field of log-based performance analysis (IC1), proposing a new approach (IC2) or extending an existing approach (IC3) for measuring process performance, and applying the presented method(s) to an industrial case study (IC4).


The final list of studies, following the above criteria, constitutes of 32 studies. For each study in the final list, standard metadata was extracted. In order to address the first research question, data about performance category, metric, and unit of measurement was extracted. For the second research question, information about what input data is required to do the analysis was extracted. Finally, for the third research question, information about tools used and underlying method was extracted. 

%% file: results.tex
\section{Results}
In this section, we examine the final list of studies from the perspective of the defined research questions. The first being the different aspects that can be assessed, followed by the data input required and available tools and methods.

\subsection{Aspects of Process Performance (\textbf{RQ1})}
The first research question concerns what aspects of process performance existing techniques measure. Not surprisingly, we found that the majority of the analyzed studies include analysis of the time perspective. This is perhaps the most basic performance aspect included in all mature performance measuring tools.

\subsubsection{Time.}
\begin{figure*}[t]
\centering
\includegraphics[scale=0.4]{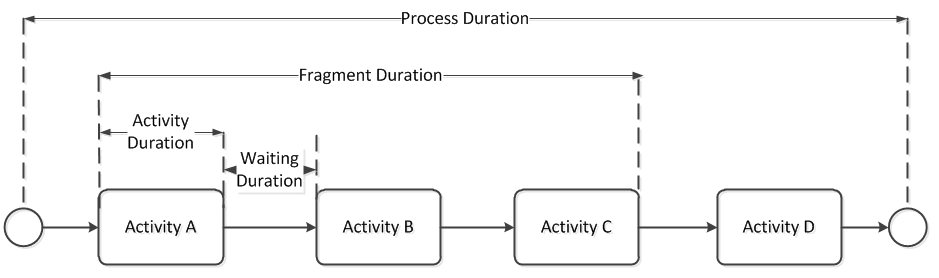}
\caption{Aspects of time performance}
\label{fig:overview}
\end{figure*}

The time aspect can be divided into four categories. These are process, fragment, activity and waiting duration (see \figurename~\ref{fig:overview}).

\paragraph{Process Duration.}
Process duration is the time distance between the start event of a process and the end event. Several techniques measure process duration. For instance, \cite{premchaiswadi2015process} examines the process duration of a peer-review process to identify bottlenecks. Similarly, Engel et al.\ \cite{engel2016analyzing} use electronic exchange messages to analyze the duration of inter-organizational processes. Different aspects of process duration can also be analyzed. For instance, \cite{leyer2011towards} examines the influence of contextual factors (such as weekday or season) on process duration. Ballambettu et al.\ \cite{ballambettu2017analyzing} propose a method for identifying key differences of process variants that could affect process duration. Suriadi et al.\, \cite{suriadi2014measuring} look at the processes of the emergency departments at four different hospitals. They compare these processes and their process duration to identify differences. Piessens et.al., \cite{piessens2010performance} recognize that some event logs contain advanced constructs such as cancelations, multiple concurrent instances, or advanced synchronization constructs. They use these constructs to gain accurate assessment of process duration.

\paragraph{Fragment Duration.}
Fragment duration considers the time required to complete a fragment (a set of activities) of a process. Wang et al.\ \cite{wang2014acquiring} propose a framework for applying process mining in logistics and analyze process fragments of a Chinese bulk port process. They identify the most time-consuming fragments of the process and, using the fragment durations, they categorize cases containing those fragment to give insights on their performance.

\paragraph{Activity Duration.}
Activity duration considers the duration of an activity. Activity duration analysis is also very common in performance analysis. For instance, \cite{cho2014systematic} applies existing process mining techniques to analyze the activity duration of a Korean hospital event log. A similar analysis is conducted by \cite{mans2008application} on a Dutch hospital log. Leyer \cite{leyer2011towards} measures the impact of contextual factors on activity duration. Activity time is analyzed in a two-step method combining process mining and statistical methods. In a similar vein, Hompes et al.\, \cite{hompes2016generic} use statistical methods to analyze the effect of context on a set of key process performance indicators at the activity level.
Activity duration analysis commonly takes an aggregated viewpoint, considering, for instance, the average time of all executed instances of each activity. However, a process have variants where activity durations vary across different variants. To address variability in activity duration based on variants, in \cite{ballambettu2017analyzing}, the authors propose a method that allows for identifying key differences of activity duration across process variants.

\paragraph{Waiting Duration.}
Waiting time in processes is one of the main wastes \cite{dumas2013fundamentals} and, as such, it has to be reduced for process improvement. As it is one of the main approaches to improve processes, waiting time analysis is the focus of many techniques. Jaisook and Premchaiswadi \cite{jaisook2015time} investigate hospital logs to examine the average duration a patient spends waiting in a private hospital. The authors do so by using the built-in functionality of Disco \cite{gunther2012disco}. Similarly, Perimal-Lewis et al.\ \cite{perimal2016application} rely on Disco to examine the processes of an emergency department. In so doing, they apply process mining to identify deviating activities in regards to waiting duration. The results highlight bottlenecks in the performance of the processes. Park et al.\, \cite{park2014integrated} propose a framework for analyzing block manufacturing processes by assessing the total waiting time. In a similar manner, in \cite{premchaiswadi2015process}, the authors examine the total waiting time of a peer-review process.
In \cite{bose2015opportunities,ballambettu2017analyzing}, the authors present a framework for analyzing similar processes across several installations. They propose a method to analyze a collection of logs from different performance perspectives, one of which is waiting duration.
An aspect of waiting duration is delay analysis. Delays refer to cases where the completion time is later than the planned completion time. Senderovich et al.\, \cite{senderovich2016conformance} analyze a process log from the perspective of operational deviations resulting in tardiness (delays) from a process duration perspective. Park et al.\, \cite{park2015workload} analyze delays in a make-to-order manufacturing firm. They define two delay indicators, activity and processing delay and found that some delays can be explained by seasonality.

\subsubsection{Resources.}
The performance of human resources is another aspect of process performance that is often analyzed. Pika et al.\, \cite{pika2014extensible} introduce an extensive framework for analysis of human resources from different perspectives. Their framework measures with the aid of time series analysis, resource utilization and productivity. Workload has been recognized as affecting resource performance as discussed in \cite{nakatumba2009analyzing}. Here, the authors explore the effect of workload on service times based on historic data and by using regression analysis. A similar metric is used by \cite{park2015workload} when analyzing manufacturing processes. In \cite{DBLP:journals/eswa/HuangLD12}, Huang at al.\ present an approach for measuring resource behavior from four perspectives, i.e., preference, availability, competence and cooperation. Resources can also be non-human such as materials. In analyzing a block manufacturing process, Park et al.\, \cite{park2014integrated} consider materials (welding length) for performance analysis.

\subsubsection{Quality.}
We also identified quality as a performance perspective. Quality can be divided into internal and external. Internal quality regards the conformance of the process outcome to internally defined targets, whereas external quality refers to customers' satisfaction with the process outcome \cite{dumas2013fundamentals}.
Internal quality analysis has been conducted by Arpasat et al.\ \cite{arpasat2015improvement} who analyze the reasons why too many attempts were required to solve a problem in a bank customer-service process. They apply Disco to identify the causes for inappropriate (not successful) interventions. External quality can be based on the analysis of the complaints received.  Wongvigran and Premchaiswadi \cite{wongvigran2015analysis} analyze a call-center log by considering the number of complaints to identify teams receiving most complaints.

\subsection{What Input Data (\textbf{RQ2})}
Automated performance analysis requires logs capturing executed events. In order to apply process mining techniques, the minimum requirement on data captured by event logs are ``case id'', ``activity'', and ``timestamp'' \cite{van2016process}. However, for performance analysis, additional data is required depending on what kind of analysis is to be conducted. The time performance of a process, be it the process, waiting, activity, or fragment duration, is commonly measured as maximum, minimum, mean, or average duration \cite{premchaiswadi2015process,piessens2010performance}.
A mere timestamp is sufficient when considering process or fragment duration as the duration is calculated by using the timestamps of the first and the last activity (of the process or fragment). However, if activity and waiting duration are to be measured, it is necessary to have the start and end time (timestamp) of each activity in the log \cite{cho2014systematic,mans2008application,hompes2016generic,bose2015opportunities}. Indeed, logs might only include one timestamp such as for activity completion. Activity and waiting duration analysis is not possible on such logs. However, in these cases, it is possible to estimate average activity and waiting times using probabilistic methods \cite{nogayama2015estimation}.

Resources are measured by considering the ``performer'' representing the human resource and/or ``materials''. Materials refer to the amount of a particular type of materials used for performing an activity. For such purposes, the input log must hold data on performers and/or amounts (quantity or costs). 
Semi-structured business processes (where the execution of the process is not fully supported by a system) do not capture all interactions among actors (e.g., interactions with customers). Logs from such processes capture the information partially. Wombacher and Lacob \cite{wombacher2013start} propose an approach to make such logs suitable for performance analysis.

Quality is either ``internally'' or ``externally'' induced. Internally induced measures commonly include binary categorization of the process outcome (desired or undesired) such as defects, errors, or delays \cite{arpasat2015improvement}. Externally induced performance refers to the determination of quality from sources external to the process such as customer complaints \cite{wongvigran2015analysis}. The log must contain data that clearly marks each case with information about internally or externally induced measures. If the data exists outside the log, a pre-processing is required to enrich the log with the required quality attributes.

\begin{table}[t!]
	\centering
\caption{Input data required for performance analysis}	
	\label{table:input}
\resizebox{\textwidth}{!}{%
\begin{tiny}
\begin{tabular}{|l|c|c|c|c|c|c|c|c|}
\hline
\multirow{2}{*}{\textbf{Attribute}} & \multicolumn{4}{c|}{\textbf{Time}} & \multicolumn{2}{c|}{\textbf{Resources}} & \multicolumn{2}{c|}{\textbf{Quality}} \\ \cline{2-9}
 & \textbf{Process} & \textbf{Fragment} & \textbf{Activity} & \textbf{Waiting} & \textbf{Performer} & \textbf{Materials} & \textbf{Internal} & \textbf{External} \\ \hline
Case Id & Y & Y & Y & Y & Y & Y & Y & Y \\ \hline
Activity & Y & Y & Y & Y & Y & Y & Y & Y \\ \hline
Timestamp & Y & Y & Y & Y & Y & Y & Y & Y \\ \hline
\begin{tabular}[l]{@{}l@{}}Activity \\ Start Time\end{tabular} & - & - & Y & Y & O & O & - & - \\ \hline
\begin{tabular}[l]{@{}l@{}}Activity \\ End Time\end{tabular} & - & - & Y & Y & O & O & - & - \\ \hline
Quality Tag & - & - & - & - & - & - & Y & Y \\ \hline
Performer & - & - & - & - & Y & O & - & - \\ \hline
Materials & - & - & - & - & O & Y & - & - \\ \hline
\end{tabular}%
\end{tiny}
}
\end{table}

\tablename~\ref{table:input} depicts the data requirements (Y) and optional requirements (O) for process performance analysis. As can be seen, the minimum requirements are case id, activity, and timestamp. It follows naturally that the more data the log holds, the more advanced performance analysis can be made.

\subsection{Approaches and Tools (\textbf{RQ3})}
There is a range of techniques to extract and analyze process performance characteristics (incl.\ performance measures) from event logs. For example, de
Leoni et al.\ \cite{DBLP:conf/bpm/LeoniAD14,DBLP:journals/is/LeoniAD16} propose a framework to extract process performance characteristics from event logs and to correlate them in order to discriminate, for example, between the performance of cases that lead to ``positive'' outcomes versus ``negative'' outcomes.
In \cite{pika2014extensible}, the authors present an extensible framework for extracting knowledge from event logs about the behavior of a human resource and for analyzing the dynamics of this behavior over time.


Another group of works is aimed at understanding the influence of contextual factors on process performance. For example, in \cite{hompes2016generic}, the authors introduce a generic context-aware analysis framework that analyzes activity durations using multiple perspectives. In \cite{DBLP:conf/bpm/ReijersSJ07}, Reijers et al.\ investigate whether the place where an actor works affects the performance of a business process. In \cite{leyer2011towards}, the authors present a methodological approach to identify the effect of contextual factors on business process performance in terms of processing time combining process mining techniques with statistical methods. This approach facilitates detecting impacted activities thus determining which activities within a business process are indeed dependent on the context. Close to the above studies are the ones presented in \cite{bose2015opportunities,ballambettu2017analyzing}. Here, starting from the observation that an organization might perform well for some clients and perform below par on others, the authors present a framework for analyzing operational event data of related processes across different clients to gain insights on process performance.



Another group of approaches is related to performance in collaborative processes. For example, in \cite{engel2016analyzing}, the authors present the EDImine Framework for enabling the application of process mining techniques in the field of EDI-supported inter-organizational business processes, and for supporting inter-organizational performance evaluation using business information from EDI messages, event logs, and process models.
In \cite{DBLP:journals/dke/HachichaFMO16}, Hachicha et al.\ present an analysis and assessment approach for collaborative business
processes in SOA in order to maintain their performance in competitive markets.

Process performance has also been approached from the perspective of queuing theory. Senderovich et al. \cite{DBLP:conf/caise/SenderovichWGM14,DBLP:journals/is/SenderovichWGM15} propose a method to discover characteristics of ``work queues'' from event logs at the level of an entire process or of individual activities. In \cite{senderovich2016conformance}, the authors target the analysis of resource-driven processes based on
event logs. In particular, they focus on processes for which there exists a predefined assignment of activity instances to resources that execute activities. The goal is to decrease the tardiness and lower the flow time.

More advanced performance analysis techniques have been recently presented in \cite{DBLP:conf/caise/NguyenDHRM16,DBLP:journals/dss/SuriadiOAH15}. In \cite{DBLP:conf/caise/NguyenDHRM16}, the authors present a technique to understand how bottlenecks form and dissolve over time via the notion of Staged Process Flow.
In \cite{DBLP:journals/dss/SuriadiOAH15}, Suriadi et al.\ present a framework based on the concept of event interval. The framework allows for a systematic approach to sophisticated performance-related analysis (e.g., resource productivity trends, recurring working patterns of resources,
waiting time distributions over time, and resource performance
comparison), even with information-poor event logs.

Other studies overlay the performance measures on top of a process model by replaying the log on the process model \cite{DBLP:journals/widm/AalstAD12,piessens2010performance} and calculating aggregate performance measures for each element in the process model during the replay.
Techniques for enhancing the quality of performance analysis based on log replay have been proposed \cite{DBLP:conf/bpm/DongenA09}.
A related technique supported by contemporary performance analysis tools is log animation. Log animation displays in a movie-like fashion how cases circulate through the
process model over time \cite{DBLP:conf/bpm/DongenA09,DBLP:conf/bpm/GuntherA07,DBLP:conf/bpm/ConfortiDRMNOR15}.

The analyzed studies mainly use ProM (15 studies) and Disco (6 studies). In the remaining studies, the authors developed their own applications.

%% file: framework.tex
\section{Framework}
In this section, we synthesize the above results in a framework aimed at assisting businesses to find the most suitable approach for performance analysis. Businesses, often not acquainted to the academic domain within this field, might find it challenging to navigate through the studies. As such, our framework might help in identifying the first steps.

\begin{table}[t!]
\centering
\caption{Framework}
\label{fram}
\resizebox{\textwidth}{!}{%
\begin{tabular}{|c|c|c|c|c|}
\hline
\multirow{2}{*}{\textbf{Type}} & \multirow{2}{*}{\textbf{Input}} & \multicolumn{3}{c|}{\textbf{Performance Perspective}} \\ \cline{3-5}
 &  & \textbf{Time} & \textbf{Resources} & \textbf{Quality} \\ \hline
\multirow{15}{*}{\begin{tabular}[c]{@{}c@{}}Descriptive\\ Performance\\ Analysis\end{tabular}} & \multirow{4}{*}{Minimum Required Data} & \begin{tabular}[c]{@{}c@{}}\cite{premchaiswadi2015process},\cite{leyer2011towards},\cite{ballambettu2017analyzing},\cite{bose2015opportunities},\cite{piessens2010performance},\cite{DBLP:journals/widm/AalstAD12}\\ Process Duration\end{tabular} & \multirow{2}{*}{-} & \multirow{2}{*}{-} \\ \cline{3-3}
 &  & \begin{tabular}[c]{@{}c@{}}\cite{wang2014acquiring}\\ Fragment Duration\end{tabular} &  &  \\ \cline{2-5}
 & \multirow{6}{*}{Activity Start and End Time} & \begin{tabular}[c]{@{}c@{}}\cite{cho2014systematic},\cite{mans2008application},\cite{leyer2011towards},\cite{hompes2016generic},\cite{ballambettu2017analyzing}\\ Activity Duration\end{tabular} & \multirow{3}{*}{-} & \multirow{3}{*}{-} \\ \cline{3-3}
 &  & \begin{tabular}[c]{@{}c@{}}\cite{jaisook2015time},\cite{perimal2016application},\cite{park2014integrated},\cite{premchaiswadi2015process},\cite{cho2014systematic},\cite{mans2008application}\\ Waiting Duration\end{tabular} &  &  \\ \cline{3-3}
 &  & \begin{tabular}[c]{@{}c@{}}\cite{park2015workload},\cite{senderovich2016conformance},\cite{DBLP:conf/caise/SenderovichWGM14},\cite{DBLP:journals/is/SenderovichWGM15}\\ Delay Duration\end{tabular} &  &  \\ \cline{2-5}
 & Internal Quality & . & - & \cite{arpasat2015improvement} \\ \cline{2-5}
 & External Quality & - & - & \cite{wongvigran2015analysis} \\ \cline{2-5}
 & Human Resources & - & \cite{pika2014extensible},\cite{nakatumba2009analyzing},\cite{park2015workload},\cite{DBLP:journals/eswa/HuangLD12},\cite{DBLP:conf/bpm/ReijersSJ07} & - \\ \cline{2-5}
 & Materials & - & \cite{park2014integrated} & - \\ \hline
\textbf{Type} & \multicolumn{4}{c|}{\textbf{Description}}  \\ \hline
\multirow{10}{*}{\begin{tabular}[c]{@{}c@{}}Complex \\ Performance\\ Analysis\end{tabular}} & \multicolumn{4}{|c|}{\begin{tabular}[c]{@{}c@{}} \cite{DBLP:conf/bpm/LeoniAD14},\cite{DBLP:journals/is/LeoniAD16}  \\ Framework to extract process characteristics from event logs discriminating between positive and negative cases\end{tabular}}  \\ \cline{2-5}
 & \multicolumn{4}{|c|}{\begin{tabular}[c]{@{}c@{}}\cite{ballambettu2017analyzing},\cite{bose2015opportunities}\\ Comparing waiting duration of similar process in different installations\end{tabular}} \\ \cline{2-5}
 & \multicolumn{4}{|c|}{\begin{tabular}[c]{@{}c@{}}\cite{engel2016analyzing},\cite{DBLP:journals/dke/HachichaFMO16}\\ Collaborative Processes\end{tabular}} \\ \cline{2-5}
 & \multicolumn{4}{|c|}{\begin{tabular}[c]{@{}c@{}}\cite{DBLP:conf/caise/NguyenDHRM16}\\ Evolution of performance over time\end{tabular} }  \\ \cline{2-5}
 & \multicolumn{4}{|c|}{\begin{tabular}[c]{@{}c@{}}\cite{DBLP:journals/dss/SuriadiOAH15}\\ Framework for performance-related analysis with information-poor event logs\end{tabular}} \\ \hline
 \textbf{Type} & \multicolumn{4}{c|}{\textbf{Domain}}  \\ \hline
 \multirow{10}{*}{\begin{tabular}[c]{@{}c@{}}Case \\ Study\end{tabular}} & \multicolumn{4}{|c|}{\begin{tabular}[c]{@{}c@{}}\cite{pika2014extensible},\cite{hompes2016generic},\cite{leyer2011towards}\\ Banks\end{tabular}}  \\ \cline{2-5}
 & \multicolumn{4}{|c|}{\begin{tabular}[c]{@{}c@{}}\cite{cho2014systematic},\cite{mans2008application},\cite{suriadi2014measuring},\cite{yampaka2016application},\cite{jaisook2015time},\cite{7975250}\\ Healthcare Processes\end{tabular}} \\ \cline{2-5}
 & \multicolumn{4}{|c|}{\begin{tabular}[c]{@{}c@{}}\cite{park2014integrated},\cite{park2015workload},\cite{engel2016analyzing}\\ Manufacturing Processes\end{tabular}}  \\ \cline{2-5}
 & \multicolumn{4}{|c|}{\begin{tabular}[c]{@{}c@{}}\cite{wang2014acquiring}\\ Logistics\end{tabular}}   \\ \cline{2-5}
 & \multicolumn{4}{|c|}{\begin{tabular}[c]{@{}c@{}}\cite{nakatumba2009analyzing},\cite{arpasat2015improvement},\cite{wongvigran2015analysis},\cite{premchaiswadi2015process}\\ Service Processes\end{tabular} } \\ \hline
\end{tabular}%
}
\end{table}

The framework considers three types of techniques. Most studies aim at descriptive performance analysis of a single log. Concerning this type of techniques, we consider two aspects. The first aspect is derived from the first research question about performance perspectives. As such, the performance perspectives are time, resources, and quality. The second aspect refers to the data available in the input logs. Depending on what data is available, different types of the performance can be analyzed. Note that the log must include at least case id, activity, and timestamp (minimum required data). Some studies compare logs of similar processes from several sites or use logs pertaining to collaborative processes. Such approaches are more complex but might be highly relevant for some businesses. Finally, we noted case studies contextualizing performance analysis to a certain domain. As such case studies are also valuable for businesses, we include them in the framework. When combining all these techniques, we gain a framework as shown in \tablename~\ref{fram}.

A business seeking to conduct data-driven performance analysis, should first select the type of technique. Descriptive analysis will show the current state and highlight cases and/or areas in the process where there are opportunities for improvement. For descriptive analysis, the minimum requirement is an input log capturing mandatory data (case id, activity, and timestamp). With this data, it is possible to perform process and fragment duration analysis. If the log contains timestamps for start and end of activities, it will be possible to conduct activity and waiting duration analysis. For delay analysis, it might be required to have scheduling data. For human resource performance analysis, the log must contain data about who performed which activity. However, resource does not need only to be human. For non-human resource analysis, the log must clearly show how much of the materials was used for each case or activity.
For quality analysis, the log must also contain, for each case, data about if the case had a desired or undesired outcome. This might be in case of a defect or complaint made.

Our review reveals approaches used for complex performance analysis.
Complex analysis covers comparative performance analysis between several installations, such as treatment processes at different hospitals or ERP systems installed at several client organizations. These approaches not only analyze the performance of each variant, but also compare them so to identify reasons for one being more efficient than the other. Another type of complex analysis is the one related to the performance of collaborative processes that can be inter- or intra- organizational. In addition, there are techniques to extract process characteristics from event logs to the aim of discriminating between positive and negative performance, techniques for the analysis of the evolution of process performance over time, and for sophisticated performance-related analysis with information-poor event logs.

Finally, the framework contains case studies from different industry domains. Most methods have validated their results on real-life industry logs and in so doing, also gained some insight that is specific for that industry. For instance, financial, healthcare, and manufacturing processes have been used to validate results. This will be valuable to businesses operating within the same industry or that have similar processes as those used for testing the results in the analyzed studies.

%% file: conclusion.tex
\section{Conclusion}
Business process performance analysis has been conducted for many decades to identify opportunities for process enhancement. In this light, performance analysis based on process mining techniques offer great value for businesses and our systematic literature review identifies tools for them to use. However, it might be difficult for businesses to navigate within this field. Therefore, we propose a framework to aid them in finding suitable methods. The framework considers the complexity of the analysis, performance perspectives, required input data, and tool availability. In particular, we show that performance is analyzed from time (process, fragment, activity, and waiting duration), resources, and quality aspects. Although process flexibility is also a performance indicator \cite{dumas2013fundamentals}, currently there are no approaches for its analysis. Therefore, an important avenue for future work in the process performance analysis field is the development of techniques for analyzing this performance perspective. 